\documentclass[a4paper,twoside,final,12pt,titlepage]{article}%
\usepackage{amsmath}
\usepackage{amsfonts}
\usepackage{amssymb}
\usepackage{graphicx}%
\providecommand{\U}[1]{\protect\rule{.1in}{.1in}}
\newtheorem{theorem}{Theorem}

\newtheorem{proposition}[theorem]{Proposition}

\newenvironment{proof}[1][Proof]{\noindent\textbf{#1.} }{\ \rule{0.5em}{0.5em}}
\newskip\humongous \humongous=0pt plus 1000pt minus 1000pt

\newif\ifdtup

\catcode`\@=11
\@addtoreset{equation}{section}

\def\@normalsize{\@setsize\normalsize{15pt}\xiipt\@xiipt
\abovedisplayskip 14pt plus3pt minus3pt\belowdisplayskip \abovedisplayskip
\abovedisplayshortskip \z@ plus3pt\belowdisplayshortskip 7pt plus3.5pt minus0pt}
\def\small{\@setsize\small{13.6pt}\xipt\@xipt
\abovedisplayskip 13pt plus3pt minus3pt\belowdisplayskip \abovedisplayskip
\abovedisplayshortskip \z@ plus3pt\belowdisplayshortskip 7pt plus3.5pt minus0pt
\def\@listi{\parsep 4.5pt plus 2pt minus 1pt
\itemsep \parsep
\topsep 9pt plus 3pt minus 3pt}}
\relax
\catcode`@=12
\catcode`\@=11
\begin{document}
\begin{titlepage}
\renewcommand{\thefootnote}{\fnsymbol{footnote}}
\bigskip
\bigskip
\bigskip
\bigskip
\bigskip
\begin{center}
{\Large  {\bf Baryons and Skyrmions in QCD \\
with Quarks in Higher Representations}
}
\end{center}
\renewcommand{\thefootnote}{\fnsymbol{footnote}}
\bigskip
\begin{center}
{\large   Stefano Bolognesi \footnote{bolognesi@nbi.dk}}
\vskip 0.20cm
\end{center}
\begin{center}
{\it      \footnotesize
The Niels Bohr Institute, Blegdamsvej 17, DK-2100 Copenhagen \O, Denmark} \\
\end {center}
\renewcommand{\thefootnote}{\arabic{footnote}}
\setcounter{footnote}{0}
\bigskip
\bigskip
\bigskip
\noindent
\begin{center} {\bf Abstract} \end{center}
We study the baryonic sector of QCD with quarks in the two index
symmetric or antisymmetric representation. The minimal gauge
invariant state that carries baryon number cannot be identified with
the Skyrmion of the low energy chiral effective Lagrangian. Mass,
statistics and baryon number do not match. We carefully investigate
the properties of the minimal baryon in the large $N$ limit and we
find that it is unstable under formation of bound states with higher
baryonic number. These states match exactly with the properties of
the Skyrmion of the effective Lagrangian. \vfill
\begin{flushleft}
December, 2006
\end{flushleft}
\end{titlepage}

\section{Introduction}

The large $N$ expansion is a major tool in the study of strongly coupled
$SU(N)$\ gauge theories \cite{'tHooft:1973jz}. In the double line notation
gluons are represented by two lines with opposite oriented arrows and quarks,
if they are in the fundamental representation, they are represented by a
single oriented line. To every Feynman diagram there corresponds a certain
topological oriented surface with a certain number of handles and holes. Holes
correspond to quark loops. Every handle suppresses the diagram by a factor
$N^{-2}$ and every hole by a factor of $N^{-1}$. The large $N$ limit is thus
dominated by diagrams with only planar gluons and fermion quantum effects are
only present in the subleading orders.

Under the\ weak\ assumption of confinement for arbitrary large $N$, a lot of
physical information can be inferred from the large $N$ limit even if the
resummation of the planar gluons is not possible \cite{Witten:1979kh}. Of
particular interest for what follows are the baryons whose gauge wave function
is%
\begin{equation}
\epsilon_{\alpha_{1}\dots\alpha_{N}}\,Q^{\alpha_{1}}\dots Q^{\alpha_{N}}
\label{wave}%
\end{equation}
It is a gauge singlet completely antisymmetric under exchange of two quarks.
The antisymmetric property of the gauge wave function (\ref{wave}) implies
that the spatial wave function is symmetric under exchange of quarks. In the
large $N$ limit the baryon can be approximated as a system of free bosons
confined in a mean potential. The mass of the Bose-Einstein condensate scales
like the number of particles $N$.\ An important property of baryons in the
large $N$ limit is that they can be identified with the solitons of the chiral
effective Lagrangian \cite{Skyrme:1961vq,Witten:1983tx}.

The large $N$ limit with a fixed number of quarks in the fundamental
representation, has the disadvantages that all the quantum corrections due to
quark loops vanishes as $\frac{1}{N}$. For example the $\eta^{\prime}$ mass
vanishes like $\frac{\mathbf{1}}{N}$ since its value comes only from the axial
$U(1)_{\mathrm{A}}$ anomaly. In order to cure these kinds of problems another
kind of limit has been suggested in the past, where the number of fundamental
quarks $N_{f}$ is send to infinity keeping fixed the ratio $\frac{N_{f}}{N}$
\cite{Veneziano:1976wm}. Although phenomenologically appealing, this kind of
limit is even more difficult to solve than the original 't Hooft limit.

Recently another kind of large $N$ limit has received considerable attention.
This is the case of quarks in the two index, symmetric or antisymmetric (S/A),
representation. Armoni, Shifman and Veneziano have discovered that a theory
with $N_{f}$ Dirac quarks in the two index S/A representation is equivalent,
in a certain bosonic subsector and in the large $N$ limit, to a theory with
$N_{f}$ Weyl quarks in the adjoint representation \cite{Armoni:2003gp}.
Particularly interesting is the antisymmetric representation since it can be
used to reproduce QCD at $N=3$.\footnote{The idea of using quarks in the two
index antisymmetric representation to reproduce QCD at $N=3$ where first
considered in \cite{Corrigan:1979xf}}. This equivalence\ becomes particularly
useful when $N_{f}=1$ since the theory with one fermion in the adjoint is
$\mathcal{N}=1$ super Yang-Mills and some non-pertubative results are known
about it. This has been used in \cite{Armoni:2003fb} to make quantitative
predictions about QCD.

Now we will face the central issue of this paper. It has been noted in
\cite{Armoni:2004uu} that, at least at a first glance, the identification
between baryons and Skyrmions in the large $N$ limit does not work. A natural
choice for the gauge wave function of the baryon is the following%
\begin{equation}
\epsilon_{\alpha_{1}\alpha_{2}\dots\alpha_{N}}\epsilon_{\beta_{1}\beta
_{2}\dots\beta_{N}}\,Q^{\alpha_{1}\beta_{1}}Q^{\alpha_{2}\beta_{2}}\dots
Q^{\alpha_{N}\beta_{N}} \label{firstguess}%
\end{equation}
where the formula holds for both the symmetric and antisymmetric
representations. This baryon is formed of $N$ quarks and so the first guess is
that its mass scales likes $N$ in the large $N$ limit. The mass of the
Skyrmion scales like $F_{\pi}^{2}$ where $F_{\pi}$ is the pion decay constant.
In the case of the quarks in higher representations $F_{\pi}$ scales like $N$,
so the mass of the Skyrmion scales like $N^{2}$ in contrast with the naive
expectation for the baryon (\ref{firstguess}). This is the puzzle we are going
to solve in this paper.

The first step towards the solution is to realize that the naive expectation
that the mass of (\ref{firstguess}) scales like $N$ is not correct. The reason
is the following. The gauge wave function (\ref{firstguess}) is symmetric
under exchange of two quarks. Since the total wave function must be
antisymmetric, this means that the space wave function must be antisymmetric
(this has also been noted in \cite{Armoni:2003jk}). The large $N$ baryon must
thus be approximated as a set free \emph{fermions} in a mean field potential.
Since fermions cannot all be in the same ground state, there is an extra term
in the energy coming from the Fermi zero temperature pressure. At this point
one could hope that this extra term could compensate the mismatch and make the
baryon mass scale like $N^{2}$. A detailed analysis shows that this is not true.

Another problem for the candidate baryon (\ref{firstguess}) comes from the
Wess-Zumino-Witten term of the effective Lagrangian for the Nambu-Goldstone
bosons \cite{Wess:1971yu,Witten:1983tw}. From this term we can read off the
statistics and the baryon number of the Skyrmion. The baryon number is
$\frac{N(N\pm1)}{2}$, where $\pm$ stands respectively for symmetric and
antisymmetric representation, and the statistics is fermionic or bosonic
accordingly if $\frac{N(N\pm1)}{2}$ is odd or even. There is no way to recover
this numbers from the baryon (\ref{firstguess}).

\qquad The topological stability of the Skyrmion in the effective Lagrangian
indicates that, at least in the large $N$ limit, there should exist a stable
state composed by $\frac{N(N\pm1)}{2}$ quarks and whose mass scales like
$N^{2}$. This is possible if there exist a color singlet wave function that
not only is composed by $\frac{N(N\pm1)}{2}$ quarks, but is also completely
antisymmetric under exchange of them. In this paper we will show that this
function exists and that $\frac{N(N\pm1)}{2}$ is the minimal amount of quarks
needed for its existence. This also confirm the stability of these baryons. In
fact any baryonic particle with a smaller number of quarks must have the extra
contributions to its mass coming from the spatial Fermi statistics.

The case of the antisymmetric representation and $N$ odd is particularly
interesting. In this case we can show that the baryon (\ref{firstguess}) is
identically zero and it is also possible to prove that any gauge invariant
quantity symmetric under exchange of two quarks vanishes identically.

The paper is organized as follows. In Section \ref{effective} we study the
effective Lagrangian and the Skyrmion properties. In Section
\ref{baryonslargeN} we study baryons at large $N$. In Section
\ref{stablebaryons} we find the stable baryons that can be identified with the
Skyrmions in the large $N$ limit. In Section \ref{moreonantisymmetric} we
consider some peculiar property of the antisymmetric representation. Finally
in Section \ref{stability}\ we discuss the stability of the Skyrmion.

\subsection{Note on conventions}

The conventions we use in the paper are the following: $Q^{\{\alpha\beta\}}$
indicates a quark in the two index symmetric representation while
$Q^{[\alpha\beta]}$ indicates a quark the two index antisymmetric
representation. When we write $Q^{\{\alpha\beta]\text{ }}$it means that the
formula we are writing is valid for both the representations. When we formally
split the quark $Q^{\{\alpha\beta]}$ into two fundamental quarks we will use
the symbols $q^{\alpha}$ and $q^{\beta}$. We reserve the name \emph{quark
}or\emph{\ higher dimensional quark }for $Q^{\{\alpha\beta]}$, while we will
use the name \emph{fundamental quarks }for $q^{\alpha}$ and $q^{\beta}$.

\section{Effective Lagrangians, Anomalies and Skyrmions \label{effective}}

Consider asymptotically free theories with $N$ colors and $N_{f}$ Dirac
fermions transforming according to the two index (anti)symmetric
representation of QCD%
\begin{equation}
\mathcal{L}=-\frac{1}{2}\mathrm{Tr~}F_{\mu\nu}F^{\mu\nu}+\sum_{k=1}^{N_{f}%
}\overline{Q}_{\{\alpha\beta]}^{k}(iD_{\mu}\gamma^{\mu}-m_{k})Q_{k}%
^{\{\alpha\beta]} \label{lagrangian}%
\end{equation}
$F_{\mu\nu}=\partial_{\mu}A_{\nu}-\partial_{\nu}A_{\mu}+ig[A_{\mu},A_{\nu}]$
is the field strength, $g$ is the coupling constant, and the covariant
derivative is $D_{\mu}=\partial_{\mu}-igA_{\mu}$. In order to have a well
defined large $N$ limit we take the product $g^{2}\,N$ to be finite.

\begin{figure}
[tbh]
\begin{center}
\includegraphics[
height=1.5947in,
width=3.9366in
]%
{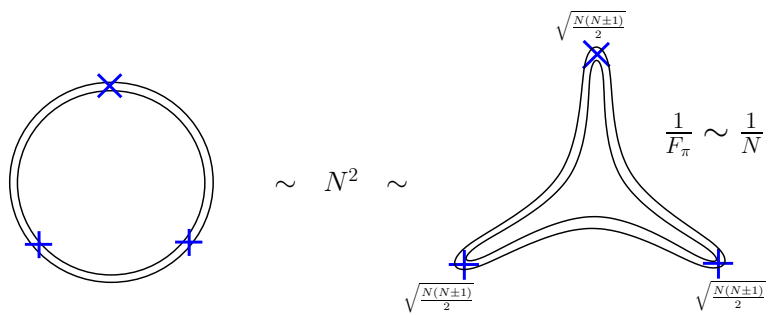}%
\caption{{\protect\footnotesize The $N$  dependence of the meson
coupling $F_{\pi}$.}}%
\label{effepai}%
\end{center}
\end{figure}

At large $N$ the theory reduces to an infinite tower of weakly coupled hadrons
whose interaction strength vanishes like $N^{-2}$. The large $N$ behavior of
these theories is similar to that of theories with fermions in the adjoint
representation of the gauge group. The dependence of the number of colors for
the meson coupling can be evaluated using the planar diagrams presented in
Figure \ref{effepai} and paying attention to the hadron wave function
normalization.
We will denote the decay constant of the typical meson by $F_{\pi}$.

\begin{figure}
[tbh]
\begin{center}
\includegraphics[
height=1.1268in,
width=3.9185in
]%
{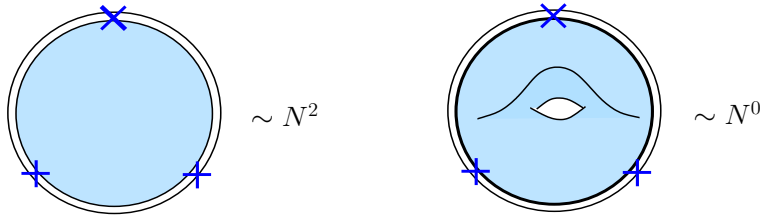}%
\caption{{\protect\footnotesize First order and second order
contributions to
the three meson interaction.}}%
\label{torus}%
\end{center}
\end{figure}

Using a double line notation the Feynman diagrams can be arranged according to
the topology of the surface related to the diagram. The $N$ powers of the
Feynman diagrams can be read off from two topological properties of the
surface: the number of handles and the number of holes. Every handle carries a
factor $N^{-2}$ and every hole carries a factor $N^{-1}$. In the ordinary 't
Hooft limit where the quarks are taken in the fundamental representation, the
holes are given by the quark loops. In the higher representation case quarks
are represented by double lines as the gluons and so there are no holes but
only handles. The contribution to $F_{\pi}$ in the large $N$ limit can thus be
arranged as in Figure \ref{torus} where the leading order is a quark closed
double line with planar quarks and gluons inside, and the next subleading
order is given by adding a handle. The leading order scales like $N^{2}$ while
the subleading order scales like $N^{0}$.
The previous color counting is not affected by the addition of a finite number
of flavors.

Here we will consider $N_{f}$ massless flavors and hence the Lagrangian has
quantum global symmetry $SU(N_{f})_{L}\times SU(N_{f})_{R}$. The global chiral
symmetry is expected to dynamically break to its maximal diagonal subgroup.
{}For the Goldstone bosons of the theory one can now construct the associated
chiral perturbation theory. The latter is an expansion in the number of
derivatives acting on the Goldstone boson fields. The low energy effective
Lagrangian describes the dynamics of the massless mesons that are the
Nambu-Goldstone bosons of the spontaneous chiral symmetry breaking. Written in
terms of the matrix $U(x)=\exp\left(  \frac{i\,\pi(x)}{F_{\pi}}\right)  $,
where $\pi(x)$ is the Goldstone boson matrix, the effective Lagrangian is%
\begin{equation}
S_{\mathrm{eff}}=\frac{1}{16}F_{\pi}^{2}\int d^{4}x\,\left\{  \mathrm{Tr}%
\partial_{\mu}U\partial_{\mu}U^{-1}+\mathrm{higher~derivatives}\right\}
+\Gamma_{\mathrm{WZW}}\mathrm{~.}%
\end{equation}
The topological term, the Wess-Zumino-Witten (WZW) term, is essential in order
to satisfy the t'Hooft anomaly conditions at the effective Lagrangian level.
Gauging the WZW term with to respect the electromagnetic interactions yields
the familiar $\pi^{0}\rightarrow2\gamma$ anomalous decay. The WZW term can be
written as
\[
\Gamma_{\mathrm{WZW}}=-i\frac{n}{240\pi^{2}}\,\int_{\mathcal{M}^{5}}%
\epsilon^{\mu\nu\rho\sigma\tau}\mathrm{Tr}\left(  \partial_{\mu}%
UU^{-1}\partial_{\nu}UU^{-1}\partial_{\rho}UU^{-1}\partial_{\sigma}%
UU^{-1}\partial_{\tau}UU^{-1}\right)  \ .
\]
where the integral must be performed over a five-dimensional manifold whose
boundary is ordinary Minkowski space. Quantum consistency of the theory
requires $n$ to be an integer.\ Matching with the underlying anomaly
computations requires $n$ to be equal to the number of quarks with respect to
the color that in the fundamental case is $N$. This imply that for the two
index representation case $n=\frac{N(N\pm1)}{2}$.

The low energy effective theory supports solitonic excitations which can be
identified with the baryonic sector of the theory. In order to obtain
classically stable configurations, it is necessary to include at least a four
derivative term in addition to the usual two derivative term. Such a term can
be for example the Skyrme term
\[
L_{\mathrm{Skyrme}}=\frac{1}{32e^{2}}\mathrm{Tr}\left(  \left[  \partial_{\mu
}UU^{-1},\partial_{\nu}UU^{-1}\right]  ^{2}\right)  \ ,
\]
but what we will say is not dependent on the details of the higher-order
terms. The Skyrmion is a texture-like solution of the effective Lagrangian
arising from the non-trivial third homotopy group of the possible
configurations of the matrix $U(x)$ (namely $\pi_{3}\left(  SU(N_{f})\right)
=\mathbf{Z}$). In the large $N$ limit we can treat the effective Lagrangian as
classical and, since the $N$ dependence appears only as a multiplicative
factor, the size and the mass of the Skyrmion scale respectively as $N^{0}$
and $\frac{N(N\pm1)}{2}$. This also implies that $e$ must scale as $1/F_{\pi}%
$. Following \cite{Witten:1983tx} we can read off the statistics and the
baryon number of the Skyrmion from the coefficient of the WZW term. The baryon
number of the Skyrmion is $\frac{N(N\pm1)}{2}$ the baryon number of the quarks
and the statistics is fermionic or bosonic accordingly if $\frac{N(N\pm1)}{2}$
is odd or even.

The results we have just obtained point all in the same direction. There
should exist in the spectrum of the theory a stable baryon that in the large
$N$ limit could be identified with the Skyrmion. This baryon should be
constituted by $\frac{N(N\pm1)}{2}$ quarks and its mass should scale like
$N^{2}$ in the large $N$ limit.

\section{Baryons at Large $N$ \label{baryonslargeN}}

\subsection{The baryon in ordinary QCD}

Now we briefly review the large $N$ behavior of the baryon in ordinary QCD.
The gauge wave function is%
\begin{equation}
\epsilon_{\alpha_{1}\dots\alpha_{N}}\,Q^{\alpha_{1}}\dots Q^{\alpha_{N}}~,
\label{ordinarybaryon}%
\end{equation}
and it is antisymmetric under exchange of two quarks.\ Since the quarks are
fermions, the total gauge function $\psi_{\mathrm{gauge}}\psi
_{\mathrm{spin/flavor}}\psi_{\mathrm{space}}$ must be antisymmetric under
exchange of two quarks. The simplest choice is to take a completely symmetric
spin wave function and a completely symmetric spatial wave function.%
\begin{equation}%
\begin{tabular}
[c]{ccc}%
$\psi_{\mathrm{gauge}}$ & $\psi_{\mathrm{spin/flavor}}$ & $\psi
_{\mathrm{space}}$\\
$-$ & $+$ & $+$%
\end{tabular}
\end{equation}

In the large $N$ limit the problem can be approximated by a system of free
bosons in a mean field potential $V_{\mathrm{mean}}\left(  r\right)  $ created
by the quarks themseves. The ground state is a Bose-Einstein condensate; the
quarks are all in the ground state of the mean field potential. The large $N$
behavior of the baryon is the following%
\begin{equation}
R\sim\mathcal{O}\left(  1\right)  ~,\qquad M\sim\mathcal{O}\left(  N\right)
~,
\end{equation}
where $R$ is the size of the baryon and $M$ its mass.

The key point to obtain this result is that the many body problem becomes
enormously simplified by the fact that the coupling constant scales like
$\frac{1}{g^{2}}\sim N$\ in the large $N$ limit. To find the mass in this many
body problem we have to sum up all the contributions from $k$-body
interactions. The $1$-body contribution is simply $N$ times the mass of the
single quark. The $2$-body interaction is of order $\frac{\mathbf{1}}{N}$ but
an additional combinatorial factor $\binom{N}{2}$ is needed and we obtain a
contribution to the energy of order $N$. In general any $k$-body interaction
is of order $\frac{\mathbf{1}}{N^{k-1}}$ in the planar limit and multiplied by
the combinatorial factor $\binom{N}{k}$ it gives a contribution of order $N$.

The same argument imply that the mean field potential $V_{\mathrm{mean}}(r)$
is constant in the large $N$ limit and so also the typical size of baryon $R$.
$R$ is in fact the with of the ground state wave function.

These arguments are consistent with the low-energy effective Lagrangian point
of view. This Lagrangian is $L_{eff}\sim N\left(  \partial U\partial
U+\partial U\partial U\partial U\partial U+\dots\right)  $ where$\ U$ is a
$SU(N_{f})$ matrix . Since $N$ is an overall multiplicative factor the radius
of the Skyrmion is of order one while its mass is of order $N$.

\subsection{The simplest baryon in higher representations}

In higher representations QCD, the simplest baryon is%
\begin{equation}
\epsilon_{\alpha_{1}\alpha_{2}\dots\alpha_{N}}\epsilon_{\beta_{1}\beta
_{2}\dots\beta_{N}}\,Q^{\{\alpha_{1}\beta_{1}]}Q^{\{\alpha_{2}\beta_{2}]}\dots
Q^{\{\alpha_{N}\beta_{N}]}~. \label{naivebaryon}%
\end{equation}
If we exchange two quarks, say for example $Q^{\{\alpha_{1}\beta_{1}]}$ and
$Q^{\{\alpha_{2}\beta_{2}]}$, this is equivalent to the exchange of
$\alpha_{1}\alpha_{2}$ in $\epsilon_{\alpha_{1}\alpha_{2}\dots\alpha_{N}}$ and
$\beta_{1}\beta_{2}$ in $\epsilon_{\beta_{1}\beta_{2}\dots\beta_{N}}$. The
result is that the gauge wave function is symmetric under exchange of two
quarks. This means that in order to have a total wave function that is
antisymmetric under exchange, the spatial wave function $\psi_{\mathrm{space}%
}$ must be antisymmetric.%
\begin{equation}%
\begin{tabular}
[c]{ccc}%
$\psi_{\mathrm{gauge}}$ & $\psi_{\mathrm{spin/flavor}}$ & $\psi
_{\mathrm{space}}$\\
$+$ & $+$ & $-$%
\end{tabular}
\end{equation}

In the large $N$ limit the problem can be approximated by a system
of free fermions in a mean field potential $V_{\mathrm{mean}}(r)$ .
The ground state is a degenerate Fermi gas and is obtained by
filling all the lowest energy states of the mean field potential up
the Fermi surface. Now there are two kind of forces that enter in
the game:
\begin{itemize}
\item[1)] Gauge forces scales like $N$ and are both repulsive and attractive,
\item[2)] Fermi zero temperature pressure scales like $N^{4/3}$ and is only
repulsive.\footnote{An important thing to note is that in the large $N$ limit
we always reach an ultrarelativistic regime due to the fact that the energy at
the Fermi surface goes to infinity as $N$ goes to infinity.}
\end{itemize}
We can thus immediately infer the following that the simplest baryon cannot be
matched with the Skyrmion; this is because the mass of the Skyrmion gous like
$N^{2}$ while the mass of this baryon obviously cannot go faster than
$N^{4/3}$.

In the next Section we shall construct the baryon that must be identifyed with
the Skyrmion of the low-energy effective Lagrangian. In Section
\ref{stability} we shall see that this baryon, although heavier than the
simplest one, is generally stable and cannot decay.

\section{The Skyrmion in the Fundamental Theory \label{stablebaryons}}

We have seen in the previous section that the simplest baryon has a gauge wave
function which is symmetric under exchange of two quarks. This has a drastic
consequence on its mass vs. $N$ dependence in the large $N$ limit. In the
following we will construct the only possible gauge wave function that is
completely antisymmetric under exchange of two quarks. We will find that the
required number of quarks, as expected from the Skyrmion analysis, is
$\frac{N(N\pm1)}{2}$.

\subsection{The symmetric representation \label{symm}}

\begin{figure}
[tbh]
\begin{center}
\includegraphics[
height=0.7057in,
width=1.4615in
]%
{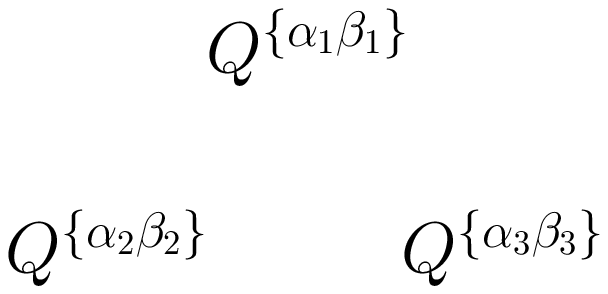}%
\caption{{\protect\footnotesize The three quarks are placed in a triangular
graphic.}}%
\label{due}%
\end{center}
\end{figure}

\begin{figure}
[tbh]
\begin{center}
\includegraphics[
height=1.606in,
width=1.2868in
]%
{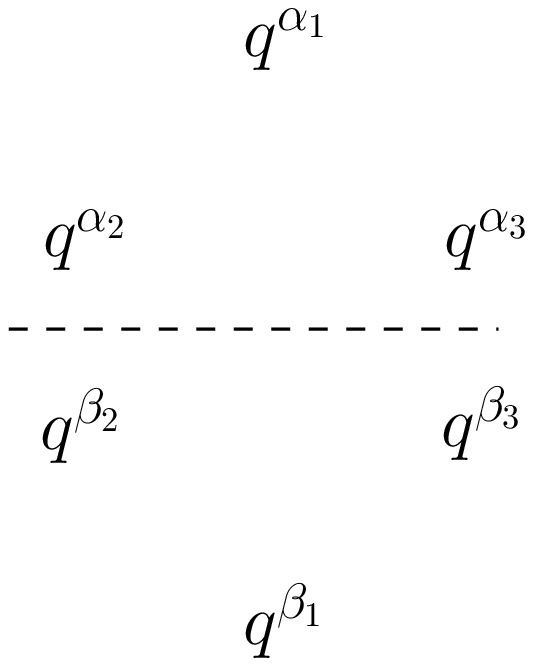}%
\caption{{\protect\footnotesize The quarks $Q^{\alpha\beta}$ are
formally splitted in two fundamental quarks $q^{\alpha}$ and
$q^{\beta}$. }}%
\label{dueraddoppiato}%
\end{center}
\end{figure}

\begin{figure}
[tbh]
\begin{center}
\includegraphics[
height=1.6008in,
width=1.6354in
]%
{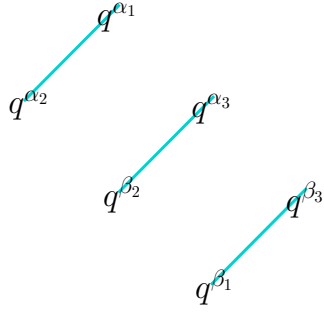}%
\caption{{\protect\footnotesize Diagrammatic representation of the
baryon for $N=2$.}}%
\label{duediagramma}%
\end{center}
\end{figure}

We start from the simplest case: $N=2$. We want to construct a gauge invariant
wave function that contains three quarks $\,Q^{\{\alpha_{1}\beta_{1}\}}$,
$Q^{\{\alpha_{2}\beta_{2}\}}$, and $Q^{\{\alpha_{3}\beta_{3}\}}$ that are
antisymmetric under exchange of two quarks. First of all we put the three
quarks in a triangular diagram like in Figure \ref{due}.%
Then we \textquotedblleft formally\textquotedblright\ split the
high-representational quarks into two fundamental quarks. For example the
quark $\,Q^{\{\alpha_{1}\beta_{1}\}}$ is splitted into two fundamental quarks
$q^{\alpha_{1}}$ and $q^{\beta_{1}}$. The diagram of Figure \ref{due} is
doubled and the result is Figure \ref{dueraddoppiato} where we have respect
the reflection symmetry with respect to the dashed line.%
Then we make a translation of the lower triangle so that the diagram becomes
that of Figure \ref{duediagramma}.%
Finally we saturate the indices with three antisymmetric tensors $\epsilon$
that are the gray lines in Figure \ref{duediagramma}. The gauge wave function
that correspond to the diagram is thus\footnote{Note also that the symmetric
representation for $SU(2)$ is equivalent to the adjoint representation and a
gauge invariant antisymmetric wave function can easily be written as
$\epsilon_{abc}Q^{a}Q^{b}Q^{c}$ where $a,b,c$ are triplet indices. This wave
function is exactly the same as that of Eq. (\ref{duesymmetricbaryon}).}%
\begin{equation}
\epsilon_{\alpha_{2}\alpha_{1}}\epsilon_{\beta_{2}\alpha_{3}}\epsilon
_{\beta_{1}\beta_{3}}\,Q^{\{\alpha_{1}\beta_{1}\}}Q^{\{\alpha_{2}\beta_{2}%
\}}Q^{\{\alpha_{3}\beta_{3}\}}~. \label{duesymmetricbaryon}%
\end{equation}
To prove that this wave function is antisymmetric under exchange of two quarks
we can proceed in two ways: algebraic and diagrammatic. The algebraic proof
is:%
\begin{align}
&  ~~~~~~\,\epsilon_{\alpha_{1}\alpha_{2}}\epsilon_{\beta_{1}\alpha_{3}%
}\epsilon_{\beta_{2}\beta_{3}}\,Q^{\{\alpha_{1}\beta_{1}\}}Q^{\{\alpha
_{2}\beta_{2}\}}Q^{\{\alpha_{3}\beta_{3}\}}\label{pedestrianproof}\\
&  =-\epsilon_{\alpha_{2}\alpha_{1}}\epsilon_{\beta_{1}\alpha_{3}}%
\epsilon_{\beta_{2}\beta_{3}}\,Q^{\{\alpha_{1}\beta_{1}\}}Q^{\{\alpha_{2}%
\beta_{2}\}}Q^{\{\alpha_{3}\beta_{3}\}}\nonumber\\
&  =-\epsilon_{\alpha_{2}\alpha_{1}}\epsilon_{\beta_{1}\beta_{3}}%
\epsilon_{\beta_{2}\alpha_{3}}\,Q^{\{\alpha_{1}\beta_{1}\}}Q^{\{\alpha
_{2}\beta_{2}\}}Q^{\{\beta_{3}\alpha_{3}\}}\nonumber\\
&  =-\epsilon_{\alpha_{2}\alpha_{1}}\epsilon_{\beta_{1}\beta_{3}}%
\epsilon_{\beta_{2}\alpha_{3}}\,Q^{\{\alpha_{1}\beta_{1}\}}Q^{\{\alpha
_{2}\beta_{2}\}}Q^{\{\alpha_{3}\beta_{3}\}}\nonumber
\end{align}
The three passages are:

\begin{itemize}
\item[(A$\rightarrow$B)] Exchange of $\alpha_{1}$ and $\alpha_{2}$ in the
$\epsilon$ that brings a minus factor,

\item[(B$\rightarrow$C)] Renomination of $\alpha_{3}$ with $\beta_{3}$ which
has no consequences,

\item[(C$\rightarrow$D)] Exchange of $\alpha_{3}$ and $\beta_{3}$ in the quark
also has no consequences.
\end{itemize}

\begin{figure}
[tbh]
\begin{center}
\includegraphics[
height=2.4613in,
width=3.7317in
]%
{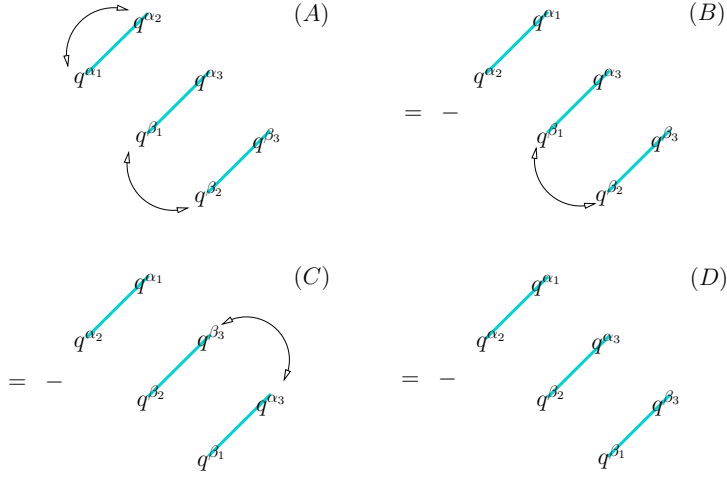}%
\caption{{\protect\footnotesize Diagrammatic proof that the baryon
of Eq. (\ref{duesymmetricbaryon}) is antisymmetric under exchange of
two quarks. The various steps $\left(A\right)$, $\left(B\right)$,
$\left(  C\right)  $ and $\left(  D\right) $ correspond to the passages of the algebraic proof (\ref{pedestrianproof}).}}%
\label{duedimostrazione}%
\end{center}
\end{figure}

The diagrammatic proof is give in Figure \ref{duedimostrazione}. Note the
correspondence between the three passages of the algebraic \ proof and the
three passages of the diagrammatic proof.%
Before going on to higher $N$ we prove a general theorem that will also give
us the recepie to build the desired gauge wave function.

\begin{proposition}
\ There is one and only one gauge wave function that is a gauge singlet and
completely antisymmetric under exchange of two quarks. This wave function is
composed by $\frac{N(N+1)}{2}$ quarks $Q^{\{\alpha\beta\}}$ and is the
completely antisymmetric subspace of the tensor product of $\frac{N(N+1)}{2}$
quarks $Q^{\{\alpha\beta\}}$.
\end{proposition}

\begin{proof}
Call $S$ the number of quarks in a hypothetical gauge wave function that
satisfies the previous conditions. We can split the quarks $Q^{\{\alpha
\beta\}}$ into $2S$ fundamental quarks $q^{\alpha_{1}}\dots q^{\alpha_{M}}$
and $q^{\beta_{1}}\dots q^{\beta_{M}}$. The wave function is the sum of
various pieces\footnote{For the wave function of $\frac{N(N+1)}{2}$ quarks
there is only one piece that corresponds to one diagram. In principle there
could be more terms added together. This in fact is the case when we consider
the antisymmetric representation.} where every piece is the partition of the
$2S$ fundamental quarks in sets of $N$ elements. These sets correspond to the
saturation with the $\epsilon$ contraction and we indicate them with a red
line that connects $N$ fundamental quarks. We need two facts to prove the
proposition: 1) $q^{\alpha_{i}}$ and $q^{\beta_{i}}$ cannot belong to the same
saturation line since the indices $\alpha_{i}$ and $\beta_{i}$, belongings to
the same quark $Q^{\{\alpha_{i}\beta_{i}\}}$, are symmetric under exchange; 2)
If $q^{\alpha_{i}}$ and $q^{\alpha_{j}}$ belong to the same saturation line,
the two partners $q^{\beta_{i}}$ and $q^{\beta_{j}}$ cannot belong to the same
saturation line . The reason is simply that an exchange of $Q^{\{\alpha
_{i}\beta_{i}\}}$ and $Q^{\{\alpha_{j}\beta_{j}\}}$ would give a plus sign
instead of the required minus sign. At this point we are ready to draw the
diagram of Figure \ref{simmetricadimostrazione}\ that is needed for the proof.
We first draw the first saturation line containing the fundamental quarks. Due
to 1) they cannot belong to the same quark and so we can call them
$q^{\alpha_{1}}\dots q^{\alpha_{N}}$. Any of these fundamental quarks must
have a partner $q^{\beta_{1}}\dots q^{\beta_{N}}$ and we draw them on the
diagram. Due to 2), any of the last quarks must belong to a different
saturation line.
\begin{figure}
[tbh]
\begin{center}
\includegraphics[
height=1.9787in,
width=1.8818in
]%
{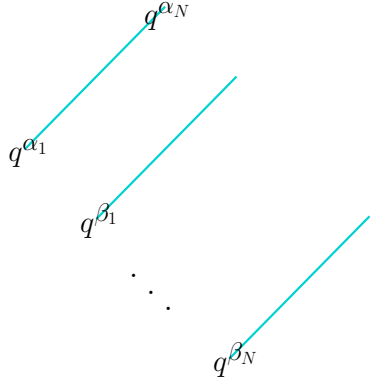}%
\caption{{\protect\footnotesize Diagrammatic proof that the minimum
number of quarks to form an antisymmetric baryon is
$\frac{N(N+1)}{2}$.}}%
\label{simmetricadimostrazione}%
\end{center}
\end{figure}
So, only starting from one saturation line, we have shown the existence of at
least $N(N+1)$ fundamental quarks in the diagram. This implies that
$S\geq\frac{N(N+1)}{2}$.

Now we need to prove the existence of this wave function. Consider the tensor
product of certain number of quarks $Q^{\{\alpha\beta\}}$. Every quark must be
considered as a vector space of dimension $\frac{N(N+1)}{2}$ over which the
group $SU(N)$ act as a linear representation. Now we take the subspace of the
tensor product that is completely antisymmetric under exchange. This subspace
is obviously closed under the action of the gauge group. If the number of
quarks is greater than $\frac{N(N+1)}{2}$ this subspace has dimension zero. If
the number of quarks is exactly $\frac{N(N+1)}{2}$ the antisymmetric subspace
has dimension one. We have thus proven that the completely antisymmetric space
of $\frac{N(N+1)}{2}$ quarks $Q^{\{\alpha\beta\}}$ is also a singlet of the
gauge group.
\end{proof}

\begin{figure}[tbh]
\begin{center}
\includegraphics[
height=2.1352in,
width=2.1387in
]{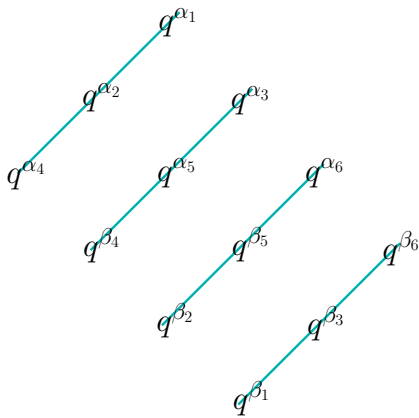}
\end{center}
\caption{The diagram for $N=3$.}%
\label{trediagramma}%
\end{figure}

The gauge wave function for general $N$ can be obtained by generalizing the
one of Figure \ref{duediagramma} for $N=2$. For example the diagram for $N=3$
is given in Figure \ref{trediagramma}. 

The baryon for $N=2$ of Figure \ref{duediagramma} does not need to be
antisymmetrized because it is already antisymmetric under exchange of any pair
of quarks. \ For $N=3$ the antisymmetrizations with respect to the four quarks
$Q^{\{\alpha_{1}\beta_{1}\}}$, $Q^{\{\alpha_{2}\beta_{2}\}}$, $Q^{\{\alpha
_{3}\beta_{3}\}}$ and $Q^{\{\alpha_{4}\beta_{4}\}}$ is enough to guarantee the
complete antisymmetrization. \ From the diagram in Figure \ref{trediagramma}
it can be seen that the antisymmetrization with respect to the exchange
$Q^{\{\alpha_{1}\beta_{1}\}}\leftrightarrow$ $Q^{\{\alpha_{2}\beta_{2}\}}$
implies that with respect to\ \ $Q^{\{\alpha_{3}\beta_{3}\}}\leftrightarrow
Q^{\{\alpha_{5}\beta_{5}\}}$ and the same for the two exchanges $Q^{\{\alpha
_{2}\beta_{2}\}}\leftrightarrow Q^{\{\alpha_{4}\beta_{4}\}}$ and
$Q^{\{\alpha_{3}\beta_{3}\}}\leftrightarrow Q^{\{\alpha_{6}\beta_{6}\}}$. We
have thus the sufficient amount of exchanges to generate the complete
permutation group.

\subsection{The antisymmetric representation \label{antisym}}

As we have done in the previous subsection we start with simplest cases and
then we try to generalize. Our goal is a gauge invariant and antisymmetric
wave function that contains $\frac{N(N-1)}{2}$ quarks $Q^{[\alpha\beta]}$. For
$N=2$, we have $\frac{N(N-1)}{2}=1$, and it is easy to find such a wave
function $\epsilon_{\gamma\delta}\,Q^{[\gamma\delta]}$. For $N=3$ we need a
wave function that contains three quarks. To guess it using directly
$Q^{[\alpha\beta]}$ is not easy, but we can use a trick. The antisymmetric
representation for $N=3$ is equivalent to the anti-fundamental $\widetilde
{Q}_{\gamma}=\frac{1}{2}\epsilon_{\gamma\alpha\beta}Q^{[\alpha\beta]}$ and we
know how to write a baryon for the anti-fundamental representation%
\begin{equation}
\epsilon^{\gamma\rho\tau}\widetilde{Q}_{\gamma}\widetilde{Q}_{\rho}%
\widetilde{Q}_{\tau}~. \label{qcd}%
\end{equation}
Substituting the relation between $\widetilde{Q}_{\gamma}$ and $Q^{[\alpha
\beta]}$ we obtain%
\begin{equation}
\frac{1}{2}(\epsilon_{\gamma_{1}\delta_{1}\alpha}\epsilon_{\gamma_{2}%
\delta_{2}\beta}-\epsilon_{\gamma_{2}\delta_{2}\alpha}\epsilon_{\gamma
_{1}\delta_{1}\beta})\,Q^{[\alpha\beta]}Q^{[\gamma_{1}\delta_{1}]}%
Q^{[\gamma_{2}\delta_{2}]}~. \label{treantisymmetricbaryon}%
\end{equation}
We know by construction that this wave function is antisymmetric under
exchange of any couple of quarks. Before going on we prove a general theorem.

\begin{proposition}
\ There is one and only one gauge wave function that is gauge singlet and
completely antisymmetric under exchange of two quarks. This wave function is
composed by $\frac{N(N-1)}{2}$ quarks $Q^{[\alpha\beta]}$ and is the
antisymmetric subspace of the tensor product of $\frac{N(N-1)}{2}$ quarks
$Q^{[\alpha\beta]}$.
\end{proposition}

\begin{proof}
Denote by $A$ the number of quarks in a hypothetical gauge wave function that
satisfies the previous conditions. The reason why $A$ can be smaller than $S$
is that now it is instead possible for a quark to have both indices on the
same saturation line. By convention we will denote these quarks by
$Q^{[\gamma\delta]}$ while the other ones, whose indices belong to different
saturation lines, will be denoted as before $Q^{[\alpha\beta]}$. Only the
quarks of the type $Q^{[\alpha\beta]}$ will be splitted into two fundamental
quarks $q^{\alpha}$ and $q^{\beta}$. For the proof we need the following two
basic facts: 1) One saturation line can contain at most one quark of the type
$Q^{[\gamma_{i}\delta_{i}]}$ otherwise the wave function will be symmetric
under exchange of these quarks. 2) If $q^{\alpha_{i}}$ and $q^{\alpha_{j}}$
belong to the same saturation line, the two partners $q^{\beta_{i}}$ and
$q^{\beta_{j}}$ cannot belong to the same saturation line . The reason is the
same as in the case of the symmetric representation. At this point we are
ready to draw the diagram of Figure \ref{antisimmetricadimstrazione}\ that is
needed for the proof. We first draw the first saturation line that contain one
quark $Q^{[\gamma_{1}\delta_{1}]}$ and $N-2$ fundamental quarks $q^{\alpha
_{1}}\dots q^{\alpha_{N-2}}$. In principle we could also have zero quarks of
the type $Q^{[\gamma\delta]}$ on one saturation line but, since we want to
minimize the number of quarks in the wave function, we will assume that every
saturation line contains one and only one quark of the type $Q^{[\gamma
\delta]}$. Due to 2), the partners of $q^{\alpha_{1}}\dots q^{\alpha_{N-2}}$,
that we denote by $q^{\beta_{1}}\dots q^{\beta_{N-2}}$, must belong to
different saturation lines and any of these lines will contain one quark
$Q^{[\gamma_{2}\delta_{2}]}\dots Q^{[\gamma_{N-1}\delta_{N-1}]}$.
\begin{figure}[tbh]
\begin{center}
\includegraphics[
height=2.1837in,
width=2.1508in
]{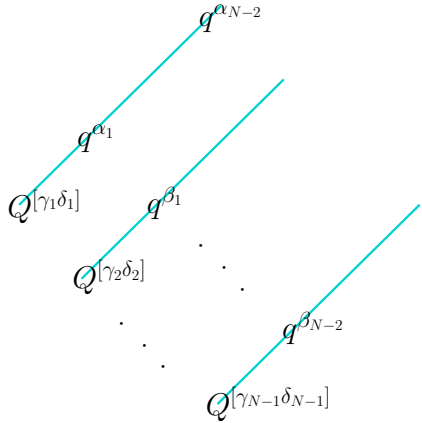}
\end{center}
\caption{{\protect\footnotesize Diagrammatic proof that the minimum
number of quarks to form an antisymmetric baryon is
$\frac{N(N-1)}{2}$.}}%
\label{antisimmetricadimstrazione}%
\end{figure}

The proof of the existence is exactly the same as that of the symmetric representation.
\end{proof}

\begin{figure}
[tbh]
\begin{center}
\includegraphics[
height=1.2324in,
width=3.8406in
]%
{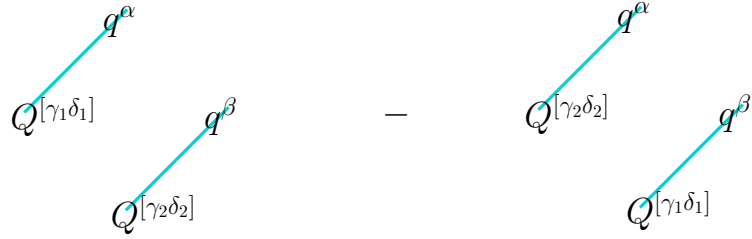}%
\caption{{\protect\footnotesize Diagrammatic representation of the baryon for
$N=3$.}}%
\label{treantidiagramma}%
\end{center}
\end{figure}

\begin{figure}[tbh]
\begin{center}
\includegraphics[
height=1.9519in,
width=1.9294in
]{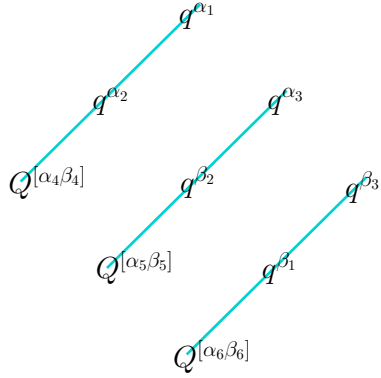}
\end{center}
\caption{{\protect\footnotesize The diagram for $N=4$.}}%
\label{quattroantidiagramma}%
\end{figure}

The wave function for $N=3$ is Eq. (\ref{treantisymmetricbaryon}) and has
already been written respecting the conventions pointed out in the proof. Now
we are ready to write the diagram. First of all there are two terms and so
there will be two diagrams. Every diagram will be divided in two parts. On one
side we put the quarks $Q^{[\gamma_{1}\delta_{1}]}$, $Q^{[\gamma_{2}\delta
_{2}]}$ and on the other side the fundamental quarks $q^{\alpha}$ and
$q^{\beta}$. \ Finally we draw the saturation lines and the two diagrams (with
the needed signs) are shown in Figure \ref{treantidiagramma}.
For example the baryon for $N=4$ is given by the diagram in Figure
\ref{quattroantidiagramma} plus the needed antisymmetrizations. For
convenience the quarks $Q^{\gamma_{1}\delta_{1}}$, $Q^{\gamma_{1,2,3}%
\delta_{1,2,3}}$ are now called $Q^{\alpha_{4,5,6}\delta_{4,5,6}}$.
The gauge wave function is
\begin{align}
&  \left(  \sum_{\sigma\in S} \mathrm{sign(\sigma)} \epsilon_{\alpha
_{\sigma(4)}\beta_{\sigma(4)}\alpha_{\sigma(2)}\alpha_{\sigma(1)}%
\epsilon_{\sigma(2)}\beta_{\sigma(5)}\beta_{\sigma(2)}\alpha_{\sigma(3)}%
}\epsilon_{\alpha_{\sigma(6)}\beta_{\sigma(6)}\beta_{\sigma(1)}\beta
_{\sigma(3)}} \right) \nonumber\\
&  Q^{\{\alpha_{1}\beta_{1}\}}Q^{\{\alpha_{2}\beta_{2}\}}Q^{\{\alpha_{3}%
\beta_{3}\}} Q^{\{\alpha_{4}\beta_{4}\}}Q^{\{\alpha_{5}\beta_{5}\}}%
Q^{\{\gamma_{3}\delta_{3}\}} \ .
\end{align}

\section{More on the Antisymmetric Representation: a Large $N$ Limit for QCD
\label{moreonantisymmetric}}

In this section we want to consider in more detail the case of the
antisymmetric representation, in particular its importance in describing a
large $N$ limit for QCD.

The simplest baryon previously considered
\begin{equation}
\epsilon_{\alpha_{1}\alpha_{2}\dots\alpha_{N}}\epsilon_{\beta_{1}\beta
_{2}\dots\beta_{N}}\,Q^{[\alpha_{1}\beta_{1}]}Q^{[\alpha_{2}\beta_{2}]}\dots
Q^{[\alpha_{N}\beta_{N}]}~, \label{minimal}%
\end{equation}
must be carefully reanalyzed in the case of the antisymmetric representation.
We have now to make a distinction between $N$ even and $N$ odd. In the case of
$N$ even (\ref{minimal}) is not the minimal baryon since we can construct a
gauge invariant wave function using only $N/2$ quarks:%
\begin{equation}
\epsilon_{\gamma_{1}\gamma_{2}\dots\gamma_{N/2}\delta_{1}\delta_{2}\dots
\delta_{N/2}}\,Q^{[\gamma_{1}\delta_{1}]}Q^{[\gamma_{2}\delta_{2}]}\dots
Q^{[\gamma_{N/2}\delta_{N/2}]}~.
\end{equation}
This baryon is symmetric under exchange of two quarks and so there is no
difference with respect to the previous conclusion: in the large $N$ limit its
mass is asymptotically proportional to $N^{4/3}$ and so it is much heavier
than the baryon constructed with $\frac{N(N-1)}{2}$ quarks.

The case of $N=2n+1$ odd is more interesting. We can prove that the minimal
baryon (\ref{minimal}) is identically zero with the following algebraic
passages:%
\begin{align}
&  ~~~~~~~~~~~~~~~~~\epsilon_{\alpha_{1}\alpha_{2}\dots\alpha_{2n+1}}%
\epsilon_{\beta_{1}\beta_{2}\dots\beta_{2n+1}}\,Q^{[\alpha_{1}\beta_{1}%
]}Q^{[\alpha_{2}\beta_{2}]}\dots Q^{[\alpha_{2n+1}\beta_{2n+1}]}\nonumber\\
&  =\,\left(  -1\right)  ^{2n+1}\epsilon_{\alpha_{1}\alpha_{2}\dots
\alpha_{2n+1}}\epsilon_{\beta_{1}\beta_{2}\dots\beta_{2n+1}}\,Q^{[\beta
_{1}\alpha_{1}]}Q^{[\beta_{2}\alpha_{2}]}\dots Q^{[\beta_{2n+1}\alpha_{2n+1}%
]}\nonumber\\
&  =~~~~~~~~~-\epsilon_{\beta_{1}\beta_{2}\dots\beta_{2n+1}}\epsilon
_{\alpha_{1}\alpha_{2}\dots\alpha_{2n+1}}Q^{[\alpha_{1}\beta_{1}]}%
Q^{[\alpha_{2}\beta_{2}]}\dots Q^{[\alpha_{2n+1}\beta_{2n+1}]}~.
\label{passages}%
\end{align}
In the first passage we have exchanged the $\alpha$ and the $\beta$ indices in
every quark. Since we have $2n+1$ quarks in the antisymmetric representation
this step brings down a minus sign. In the second step we have just renamed
$\alpha_{i}$ with $\beta_{i}$ and vice versa and this has no consequences. The
last line of (\ref{passages}) is equal to minus the fist line (a part from an
irrelevant exchange in the position of the two epsilons) and thus the wave
function must be zero. We will now prove a stronger statement:

\begin{proposition}
For $N$ odd and quarks in the antisymmetric representation, it is not possible
to write a gauge invariant wave function that is completely symmetric under
exchange of two quarks.
\end{proposition}

\begin{proof}
Take a generic wave function that is gauge invariant and symmetric under
exchange of two quarks. We are going to prove that it is identically zero.
This wave function is composed by a number of quarks that we generically
denote by $M$. $M_{\alpha\beta}$ of these quarks are of type $Q^{[\alpha
\beta]}$ and $M_{\gamma\delta}$ are of type $Q^{[\gamma\delta]}$ so that we
can write%
\begin{equation}
M=M_{\left(  \alpha\beta\right)  }+M_{\left(  \gamma\delta\right)  }~.
\label{emme}%
\end{equation}
The $M$ quarks can be divided into various \emph{connected} components, where
the connection is given by the epsilon contractions and the quarks
$Q^{[\alpha\beta]}$. Let us assume for the moment that we have only one
connected component. It is easy to see that $M_{\alpha\beta}$ must be odd. We
will now\ use the same argument we have used to show that (\ref{passages}) is
identically zero. Namely we will show that the wave function is equal to minus
itself. first we exchange all the $\alpha$ indices with their $\beta$ partners
and this contributes a minus sign since $M_{\alpha\beta}$ is odd. Then we make
a suitable number of exchange between the quarks $Q^{[\gamma\delta]}$ in order
to recover the original epsilon structure. These exchanges do not affect the
wave function since by definition it is symmetric under exchanges of two
quarks. So we have recovered the original wave function but with a minus sign
in front.

We now have to consider the more general situation in which the $M$ quarks are
divided in various disconnected components. It can easily be seen that in this
case the sub-connected components must be closed under the exchange of two
generic quarks. Put in another way, if the global wave function is symmetric
under exchange of two quarks, then also the sub-connected wave functions are
symmetric under exchange of two quarks. In this case we can thus remake the
passage of the previous paragraph but only on a sub-connected wave function
and we obtain the desired result.
\end{proof}

The previous proposition does not exclude the possible existence of a gauge
invariant wave function with less then $\frac{N(N-1)}{2}$ quarks and in a non
singlet representation of the permutation group. In this case the baryon is
not a simple product of gauge, spin and space wave function but a sum
$\sum_{i}\psi_{\mathrm{gauge}}^{i}\psi_{\mathrm{spin}}^{i}\psi_{\mathrm{space}%
}^{i}$, where $\psi_{\mathrm{gauge}}^{i}$ is the non-singlet representation of
the permutation group.

\section{Stability of the Skyrmion \label{stability}}

We want now to discuss the issue of the stability of the Skyrmion. The
Skyrmion correspond to the baryon that contains $\frac{N\left(  N\pm1\right)
}{2}$ quarks and that is fully antisymmetric in the gauge wave function. The
mass is thus proportional to the number of costituents quarks. Seen from the
low-energy effective Lagrangian, the Skyrmion is absolutly stable. In the full
theory, on the other hand, we should consider the possibility of decay into
baryons with lower numbers of costituents quarks, for example the baryon
$\epsilon_{\alpha_{1}\alpha_{2}\dots\alpha_{N}}\epsilon_{\beta_{1}\beta
_{2}\dots\beta_{N}}\,Q^{\{\alpha_{1}\beta_{1}]}Q^{\{\alpha_{2}\beta_{2}]}\dots
Q^{\{\alpha_{N}\beta_{N}]}$. This states are not visible from the low-energy
effective Lagrangian. As we have seen in Section \ref{stablebaryons} baryons
with a number of costituents quarks lower than $\frac{N\left(  N\pm1\right)
}{2}$ can not be in a fully antisymmetric gauge wave function. This imply that
the Skyrmion is the state that minimizes the mass per unit of baryon number.

\begin{figure}
[ptbh]
\begin{center}
\includegraphics[
height=1.8766in,
width=2.9793in
]%
{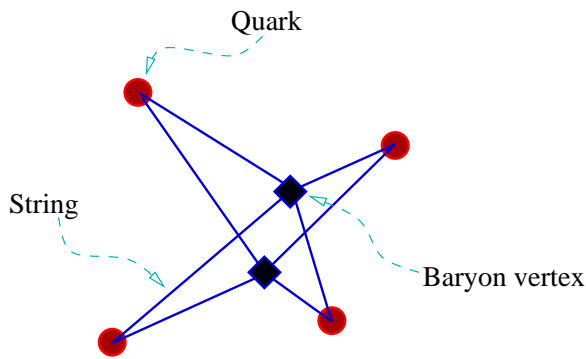}%
\caption{{\protect\footnotesize A model of the baryon (here for four colors).
Every quark is attached to two confining strings and every baryon vertex to
$N$ confining strings.}}%
\label{toymodel}%
\end{center}
\end{figure}

\begin{figure}
[ptbh]
\begin{center}
\includegraphics[
height=1.6786in,
width=3.5241in
]%
{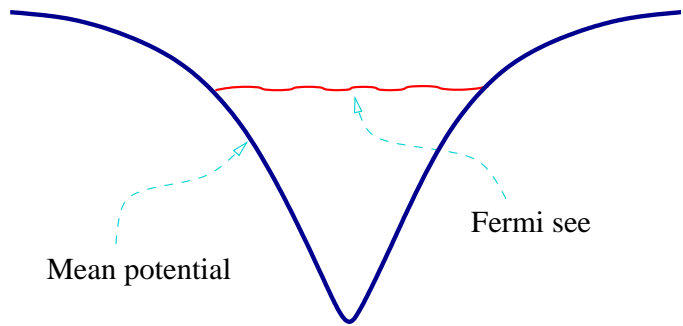}%
\caption{{\protect\footnotesize The mean potential for our toy model of the
baryon.}}%
\label{potentialbaryon}%
\end{center}
\end{figure}

Let us consider an explicit example in more detail. A Skyrmion that contains
$\frac{N\left(  N\pm1\right)  }{2}$ can decay into $\frac{N\pm1}{2}$ baryons
composed by $N$ quarks. The baryon number is conserved and so this decay
channel is in principle possiple. In order to analize the energetic of this
baryon, we propose now a toy model to schematize the fundamental baryon. We
have $N$ quarks and $2$ baryon vertices. Every quark is attached to two
fundamental strings and every baryon vertex to $N$ fundamental strings (see
Figure \ref{toymodel} for an example).
Baryon vertices have a mass of order $N$; we can thus neglect their dynamics
and consider them at rest and positioned in what we define to be the center of
the baryon. In this approximation, the quarks do not interact directly between
each other; they live in a mean potential given by the string tension
multiplyed by the distance from the center%
\begin{equation}
V_{\mathrm{mean}}(R)=2T_{\mathrm{string}}\left\vert R\right\vert
~.\label{confiningpotential}%
\end{equation}
Quarks are antysymmetric in the space wave function and so they they fill the
energy levels up to the Fermi surface (see Figure \ref{potentialbaryon}). We
indicate as $R_{\mathrm{F}}$ and $P_{\mathrm{F}}$ respectively the Fermi
radius and momentum. The total energy and the number of quarks $N$ are given
by the following integrals over the phase space:%
\begin{align}
\int^{R_{\mathrm{F}}}\int^{P_{\mathrm{F}}}\frac{d^{3}Rd^{3}P}{\left(
2\pi\right)  ^{3}}\left(  P+V_{\mathrm{mean}}(R)\right)   &  =E~,\nonumber\\
\int^{R_{\mathrm{F}}}\int^{P_{\mathrm{F}}}\frac{d^{3}Rd^{3}P}{\left(
2\pi\right)  ^{3}} &  =N~.\label{integrals}%
\end{align}
Since the quarks are massless we take the Hamiltonian to be
$P+V_{\mathrm{mean}}(R)$. From now on we neglect numerical factors such as the
phase space volume element, at this level of approximation they are not
important. The second equation of (\ref{integrals}) gives a relation between
the Fermi momentum and the Fermi radius, namely $P_{\mathrm{F}}\sim
N^{1/3}/R_{\mathrm{F}}$. The first equation of (\ref{integrals}) gives the
following expression of the energy as function of the radius
\begin{equation}
E\sim\frac{N^{4/3}}{R_{\mathrm{F}}}+T_{\mathrm{string}}NR_{\mathrm{F}}~.
\end{equation}
Minimizing we obtain $R_{\mathrm{F}}\sim N^{1/6}/\sqrt{T_{\mathrm{string}}}%
$,$\ $and\ consequently$\ P_{\mathrm{F}}\sim N^{1/6}\sqrt{T_{\mathrm{string}}%
}$. The mass of the baryon is thus given by%
\begin{equation}
M_{N-\mathrm{Baryon}}\sim N^{7/6}\sqrt{T_{\mathrm{string}}}~.
\end{equation}
The important thing to note is the $N^{7/6}$ dependence. The mass per unit of
baryon number grows as $N^{1/6}$. The Skyrmion has instead mass per unit of
baryon number of order one. This imply that the  Skyrmion is the most
convenient baryonic state since it minimizes the energy per unit of baryon number.%

This approximation breaks down when the Fermi energy $N^{1/6}\sqrt
{T_{\mathrm{string}}}$ becomes much grater than the dynamical scale. Due to
asymptotic freedom, the highly energetic quarks do not feel a confining
potential like (\ref{confiningpotential}) but instead a Coulomb-like
potential. The mass per unit of barion number stops to grow as $N^{1/6}$ and
saturates to a constant.

\section*{Acknowledgements}

I thank especially F.~Sannino for the suggestion to work on this problem and
for the many useful discussions. I thank S.~B.~Gudnason for discussions and
for the precious help with the manuscript. The present work has been presented
at the conference CAQCD in Minneapolis. I want to thank M.~Shifman and
T.~Cohen for interesting observations. This work is supported by the Marie
Curie Excellence Grant under contract MEXT-CT-2004-013510.

\end{document}